\documentclass[preprint, 12pt]{elsarticle}
\newcommand{\ket}[1]{|{#1}\rangle}
\newcommand{\bra}[1]{\langle{#1}|}

\newcommand{\be}{\begin{equation}}
\newcommand{\ee}{\end{equation}}
\newcommand{\llt}{\ll}

\usepackage{amsfonts}
\usepackage{amsmath}
\usepackage{rotating}
\usepackage{graphicx}
\usepackage{graphics}
\usepackage{epsfig}
\usepackage{amsthm}
\usepackage{amssymb}
\usepackage{amsmath}
\usepackage{amssymb}
\usepackage{graphicx}
\usepackage{lscape}
\usepackage{color}
\usepackage{epstopdf}

\journal{XX}

\begin{document}

\begin{frontmatter}

\title{Unitary fermions and L{\"u}scher's formula on a crystal}

\author{Manuel Valiente}
\address{SUPA, Institute of Photonics and Quantum Sciences, Heriot-Watt University, Edinburgh EH14 4AS, United Kingdom}
\author{Nikolaj T. Zinner}
\address{Department of Physics \& Astronomy, Aarhus University, 8000 Aarhus C, Denmark}

\begin{abstract}
We consider the low-energy particle-particle scattering properties in a periodic simple cubic crystal. In particular, we investigate the relation between the two-body scattering length and the energy shift experienced by the lowest-lying unbound state when this is placed in a periodic finite box. We introduce a continuum model for s-wave contact interactions that respects the symmetry of the Brillouin zone in its regularisation and renormalisation procedures, and corresponds to the na{\"i}ve continuum limit of the Hubbard model. The energy shifts are found to be identical to those obtained in the usual spherically symmetric renormalisation scheme upon resolving an important subtlety regarding the cutoff procedure. We then particularize to the Hubbard model, and find that for large finite lattices the results are identical to those obtained in the continuum limit. The results reported here are valid in the weak, intermediate and unitary limits, and can be used for the extraction of scattering information ,via exact diagonalisation or Monte Carlo methods, of two-body systems in realistic periodic lattices.  
\end{abstract}


\end{frontmatter}

\section{Introduction}
The notion of low-energy universality in particle-particle collisions is a powerful concept that has been around, in one form or another, for a very long time. The general idea consists of replacing realistic, complicated interactions with much simpler ones and renormalising their bare coupling constants in favour of exact (either experimental or theoretical) scattering properties. The model interaction, irrelevant at the two-body level, can then be used to investigate the effect of interactions with higher particle numbers without the extra complications of the realistic interactions. For example, in 1957, Huang and Yang introduced their s-wave regularised pseudopotential \cite{Huang}, which aimed at reproducing the exact scattering length of a realistic two-body process by means of a very simple model interaction. The idea, however, was put forward much earlier by Fermi in the 1930s \cite{Fermi}, who studied neutron-nucleon scattering, and was able to fit the scattering length in the first Born approximation.

The calculation of two-body scattering properties with rather realistic two-body interactions, if these are spin-independent, spherically symmetric, and single-channel, is not too complicated, especially with the computers we have access to nowadays. It is genuine multichannel collisions that can become quite challenging to approach directly from the corresponding Lippmann-Schwinger equation. In nuclear physics, there is a fundamental interest in obtaining nucleon-nucleon scattering properties from first principles using lattice QCD (see \cite{NuclearForces} and references therein). In condensed matter and cold atomic physics, particle-particle scattering theory in a crystal, where the incident waves are Bloch waves, and with a realistic (e.g. screened Coulomb) interaction, constitutes a formidable problem: all bands of the periodic potential are coupled by the interaction, Galilean relativity does not hold so that the collisional properties depend on the total center of mass momentum, and spherical symmetry is not present and cannot be exploited. 

In situations like those described above, it is of great appeal to be able to extract scattering properties without having to directly use scattering theory. In massive quantum field theories, L{\"u}scher showed how to extract partial-wave scattering amplitudes, i.e. phase shifts and scattering lengths, from the energy shifts due to the interactions when the system is placed in a finite volume \cite{Luscher}, thereby generalising an older result from non-relativistic quantum mechanics \cite{Huang}. However, neither L{\"u}scher's nor Huang-Yang's results for the scattering lengths hold when these are unnaturally large, i.e. near the unitary limit. This problem was elegantly solved using pionless effective field theory \cite{BeanePDS} by Beane {\it et al.} in ref. \cite{Beane2004}. In this way, an analysis of low-energy nucleon-nucleon collisions, directly from QCD, was possible soon after \cite{BeaneQCD}. In the case of particle-particle scattering in a crystal, finite-size effects in the forms given in refs. \cite{Luscher} and \cite{Beane2004}, however, do not appear to be known. In this article we derive the relevant formulas from effective field theory on a tight-binding lattice, that is, the Hubbard model. We will focus on the three-dimensional (space) case, for the one-dimensional case is well understood \cite{ValienteLattices} and L{\"u}scher's formula in one dimension \cite{Luscher} holds. We derive the pertinent expressions using Bethe-Goldstone equations, and we begin by proving these via the introduction of a more general class of methods consisting of kernel subtractions. We then introduce the Hubbard model and analyse, in first instance, its na{\"i}ve continuum limit. We then return to the Hubbard model itself and show that, at low energy and large volume, the energy shifts are completely analogous to those obtained from the na{\"i}ve continuum limit, which shows once more how low-energy physics, at least in the two-particle case, is universal.  

\section{Kernel subtractions and Bethe-Goldstone theory}
For general interactions, one way of establishing the relation between the scattering states of a system (i.e. the positive energy states in the infinite size limit) and the energy shifts in a finite box is to use Bethe-Goldstone (BG) theory \cite{FetterWalecka} in vacuum, that is, without the static Fermi sea background. In fact, even if it looks very different from it, L{\"u}scher rederived, and used the exact same old Bethe-Goldstone's equation in his seminal paper \cite{Luscher}. Here we will relate BG theory to a much more general method -- kernel subtractions -- for solving homogeneous integral equations \footnote{Here, integral equation is taken in the most general sense, and applies equally to discrete kernels.}, and show that it corresponds to a particular choice of kernel subtraction. The method is easy to implement. Let $\ket{\psi}$ be the solution to the integral equation
\begin{equation}
\ket{\psi}=\hat{K}(E)\ket{\psi},\label{homo}
\end{equation}
where $\hat{K}(E)$ is the kernel (operator), with $E$ some parameter, which in the relevant case here will correspond to an energy eigenvalue. The above equation (\ref{homo}) is homogeneous and therefore only has a solution for particular values of $E$. Consider now a vector $\ket{\bar{\psi}}$ that satisfies the following integral equation
\begin{equation}
\ket{\bar{\psi}}=\ket{F}+\left[\hat{K}(E)-\ket{F}\bra{\Gamma}\right]\ket{\bar{\psi}},
\end{equation}
where $\ket{F}$ and $\bra{\Gamma}$ are arbitrary vectors. Then, for the values of the energy for which $\langle \Gamma |\bar{\psi} \rangle =1$, it holds that $\ket{\bar{\psi}}=\ket{\psi}$.

We now particularize to the physically relevant case, and derive the BG equation. Let $\ket{\psi}$ satisfy the stationary 
Schr{\"o}dinger equation
\begin{equation}
H\ket{\psi}=E\ket{\psi}.
\end{equation}
We assume the Hamiltonian is of the usual form $H=H_0+V$. Then, if $E$ is not in the spectrum of $H_0$, the wave function satisfies the homogeneous Lippmann-Schwinger equation
\begin{equation}
\ket{\psi}=G_0(E)V\ket{\psi},
\end{equation}
where $G_0(E)=(E-H_0)^{-1}$. The kernel, Eq. (\ref{homo}), is therefore given by $\hat{K}(E)=G_0(E)V$. We choose the following vectors $\ket{F}$ and $\ket{\Gamma}$,
\begin{align}
\ket{F}&=\ket{\mathbf{k}},\\
\bra{\Gamma}&=\bra{\mathbf{k}}G_0(E)V,
\end{align}
where $\ket{\mathbf{k}}$ is an eigenstate of $H_0$, i.e. $H_0\ket{\mathbf{k}}=\epsilon(\mathbf{k})\ket{\mathbf{k}}$. Then, $\ket{\psi}$ satisfies 
\begin{align}
\ket{\psi}&=\ket{\mathbf{k}}+\left[G_0(E)V-\ket{\mathbf{k}}\bra{\mathbf{k}}G_0(E)V\right]\ket{\psi},\label{BG1}\\
E&=\epsilon(\mathbf{k})+\langle\mathbf{k}|V|\psi\rangle.\label{BG2}
\end{align}
Eq. (\ref{BG1}) together with the auxiliary condition (\ref{BG2}) correspond exactly to BG theory in operator form. In order to relate it to the finite-size reaction matrix, we introduce the identity $1=\sum_{\mathbf{k}}\ket{\mathbf{k}}\bra{\mathbf{k}}$ and define $V\ket{\psi}\equiv \hat{r}\ket{\mathbf{k}}$. After straightforward algebra, we find
\begin{align}
\langle \mathbf{k}' |\hat{r}|\mathbf{k}\rangle&=\langle \mathbf{k}' |V|\mathbf{k}\rangle + \sum_{\mathbf{q}\ne \mathbf{k}} \frac{\langle\mathbf{k}'|V|\mathbf{q}\rangle \langle \mathbf{q} | \hat{r} | \mathbf{k}\rangle}{E-\epsilon(\mathbf{q})},\label{discreteRmatrix}\\
E&=\epsilon(\mathbf{k})+\langle \mathbf{k} |\hat{r} | \mathbf{k} \rangle.\label{BG3}
\end{align}

\section{Hubbard model}
We shall focus on one specific model, namely the Hubbard model on a simple cubic lattice. This is the minimal model for interacting electrons (or any other kind of spin-$1/2$ fermions) in a crystal, and consists of a nearest-neighbour hopping, or kinetic energy term, and an on-site interaction term. Its Hamiltonian in second quantised form is given by
\begin{equation}
H=-J\sum_{\langle i,j\rangle,\sigma=\uparrow,\downarrow} c_{i\sigma}^{\dagger}c_{j\sigma}+U\sum_{i}n_{i\uparrow}n_{i\downarrow}+6JN.\label{HubbardHamiltonian}
\end{equation}
Above, $J>0$ is the hopping rate, $\langle i,j\rangle$ denotes nearest neighbours, with $i,j\in \mathbb{Z}^3$, $c_{i\sigma}$ ($c_{i\sigma}^{\dagger}$) are fermionic creation and annihilation operators, $U$ is the on-site interaction strength, $n_{i\sigma}=c_{i\sigma}^{\dagger}c_{i\sigma}$ is the local number operator and $N=\sum_{i\sigma}n_{i\sigma}$ is the total number operator (we have set the lattice spacing $d\equiv 1$). In Hamitonian (\ref{HubbardHamiltonian}), the last term on the right-hand side amounts to an energy shift so that the single-particle ground state energy vanishes. In particular, the kinetic energy dispersion, $\epsilon(\mathbf{k})$ of the Hubbard model (\ref{HubbardHamiltonian}) is given by
\begin{equation}
\epsilon(\mathbf{k})=-2J\sum_{\alpha=x,y,z}\left[\cos(k_{\alpha})-1\right],\label{dispersionHubbard}
\end{equation}
where the quasi-momenta $k_{\alpha}\in [-\pi,\pi)$, and the Brillouin zone is given by $\mathrm{BZ}\equiv [-\pi,\pi)^3$. If the system is placed on a finite cube of side length $L=L_s$ (recall the lattice spacing $d\equiv 1$)  with periodic boundary conditions, the quasi-momenta are restricted to take values $k_{\alpha}=2\pi n_{\alpha}/L$ (mod $2\pi$), with $n_{\alpha}\in \{0,1,\ldots,L-1\}$. 

\section{Na{\"i}ve continuum limit}
Before studying the Hubbard model in detail, it is very instructive to consider its na{\"i}ve continuum limit. The bare continuum Hamiltonian is given by
\begin{equation}
H=\sum_{\sigma=\uparrow,\downarrow}\frac{\hbar^2}{2m} \int_{\mathbf{r}\in [0,L)^3}d\mathbf{r} \nabla \psi^{\dagger}_{\sigma}(\mathbf{r})\cdot \nabla \psi_{\sigma}(\mathbf{r}) + g \int_{\mathbf{r}\in [0,L)^3} d\mathbf{r} \psi^{\dagger}_{\uparrow}(\mathbf{r})\psi^{\dagger}_{\downarrow}(\mathbf{r})\psi_{\downarrow}(\mathbf{r})\psi_{\uparrow}(\mathbf{r}).\label{ContinuumHamiltonian}
\end{equation}
Above, $m$ is the effective mass, $g$ is the bare interaction strength, and $\psi_{\sigma}$ ($\psi_{\sigma}^{\dagger}$) is the fermionic annihilation (creation) operator in the continuum. 

The above Hamiltonian, Eq. (\ref{ContinuumHamiltonian}), is obtained as follows. Firstly, we restore the lattice spacing $d$ (which was set to $d=1$ in the previous section), in such a way that the action of the non-interacting Hamiltonian on a single-particle wave function, in the first quantisation, is given by
\begin{equation}
(H_0\psi)(\mathbf{r})=-J\sum_{\alpha=x,y,z} \left[\psi(\mathbf{r}+\hat{e}_{\alpha} d)+\psi(\mathbf{r}-\hat{e}_{\alpha}d)-2\psi(\mathbf{r})\right],\label{H0psi}
\end{equation}
where $\mathbf{r}=(x,y,z)=(n_x,n_y,n_z)d$, with $n_{\alpha}\in \mathbb{Z}$, and where we have defined the unit vectors $\hat{e}_{\alpha}$ such that $\hat{e}_{\alpha}\cdot \mathbf{r} = \alpha$ ($\alpha=x,y,z$). The continuum limit of Eq. (\ref{H0psi}) is attained by setting $J=\hbar^2/2md^2$. The continuum limit of the bare interaction strength is attained by identifying $U=g/d^3$ as $d\to 0$. 

The effective interaction in Eq. (\ref{ContinuumHamiltonian}) is a zero-range "s-wave" interaction. It is renormalisable, and accounts correctly for the s-wave scattering length with a vanishing effective range. This interaction can be renormalised in a number of ways, all equivalent to each other. For instance, minimal subtraction in the hard cutoff regularisation scheme \cite{BeanePDS}, dimensional regularisation, Huang's pseudopotential \cite{Huang}, or the more recent method of Tan's distributions \cite{Tan,ValienteTan}, all give the same results. In all the aforementioned methods, spherical symmetry of the contact interaction is exploited. The question arises of whether the cubic symmetry of the lattice Brillouin zone in the continuum limit changes the status of affairs. The answer to this question, which seems quite trivial, is not straightforward at all. On the one hand, the scattering properties at low energies remain unchanged. On the other hand, L{\"u}scher's formula heavily relies on spherical symmetry, a fact which is particularly evident in the pionless EFT considered by Beane {\it et al.} in ref. \cite{Beane2004}, and we will show that L{\"u}scher's formula {\it does} change significantly when the lattice symmetry is respected in the regularisation process, unless special care is taken in the renormalisation process.

\subsection{Scattering theory}
In this subsection, we shall obtain the reaction matrix ($R$-matrix), by regularising the contact interaction in such a way that the symmetry of the Brillouin zone is manifestly preserved at all steps. The regularised contact interaction $V(\mathbf{k},\mathbf{k}')$ in the momentum representation for a spin-singlet fermion pair with ultraviolet (UV) cutoff $\Lambda$ is therefore given by
\begin{equation}
\langle\mathbf{k}'|V|\mathbf{k}\rangle \equiv V(\mathbf{k},\mathbf{k}') = g \theta(\Lambda,\mathbf{k}')\theta(\Lambda,\mathbf{k}),\label{intcube}
\end{equation}
where we have defined 
\begin{equation}
\theta(\Lambda,\mathbf{k}) = \prod_{\alpha=x,y,z}\theta(\Lambda^2-k_{\alpha}^2).
\end{equation}
Above, $\theta(q)$ is the Heaviside step function. The above interaction, Eq. (\ref{intcube}), has to be contrasted with the widely used spherically symmetric regularisation \cite{BeanePDS,Beane2004,ValienteTan}, which reads $g\theta(\Lambda^2-\mathbf{k}^2)$.

After separation of center of mass and relative coordinates, and at vanishing center of mass momentum, the bare $R$-matrix $\hat{R}_*$with the interaction in Eq. (\ref{intcube}) is straightforward to calculate, and reads
\begin{equation}
\langle \mathbf{k}'|\hat{R}_*(E)|\mathbf{k}\rangle = \frac{\theta(\Lambda,\mathbf{k}')\theta(\Lambda,\mathbf{k})}{1/g+I_0(E)/(2\pi)^3},\label{bareRmatrix}
\end{equation}
where 
\begin{equation}
I_0(E)=\mathcal{P}\int_{\mathbf{q}\in [-\Lambda,\Lambda)^3}d\mathbf{q} \frac{1}{E-\hbar^2q^2/m},\label{I0}
\end{equation}
with $\mathcal{P}$ denoting Cauchy's principal value. The UV structure of $I_0(E)$, Eq. (\ref{bareRmatrix}), can be extracted from its value at zero energy. Numerically, we have obtained $I_0(0)=-15.3482484734169\ldots \Lambda+\mathcal{O}(1/\Lambda)$. The zero-energy on-shell $R$-matrix can be renormalised in favour of the scattering length $a$ by means of minimal subtraction, i.e. by setting
\begin{equation}
\frac{1}{g}=\frac{1}{g_R}-\frac{I_0(0)}{(2\pi)^3},\label{Renorm1}
\end{equation}
in which case the renormalised coupling constant takes the value $g_R=4\pi \hbar^2 a/m$. The $R$-matrix at arbitrary energy is renormalisable if $I_0(E)-I_0(0)$ is finite for all energies $E$, that is, if the linearly divergent part of the integrals $I_0(E)$ is energy-independent. To show that this is indeed the case, write the integral $I_0(E)$ as
\begin{equation}
I_0(E)=\mathcal{P}\int_{q<\Lambda}d\mathbf{q} \frac{1}{E-\hbar^2q^2/m} + \int_{\mathbf{q}\in [-\Lambda,\Lambda)^3}d\mathbf{q}\frac{\theta(q^2-\Lambda^2)}{E-\hbar^2q^2/m}.\label{I0sep}
\end{equation}
The second integral on the right hand side of Eq. (\ref{I0sep}) can be calculated as a series expansion that is convergent for any finite energy. We expand the integrand as
\begin{equation}
\frac{\theta(q^2-\Lambda^2)}{E-\hbar^2q^2/m}=-\frac{\theta(q^2-\Lambda^2)}{\hbar^2q^2/m}\sum_{n=0}^{\infty} \left(\frac{mE}{\hbar^2q^2}\right)^n.
\end{equation}
The resulting integral in Eq. (\ref{I0sep}) is therefore given by
\begin{equation}
I_0(E)=I_0(0)-\frac{m}{\hbar^2}\sum_{n=1}^{\infty}\int_{\mathbf{q}\in [-\Lambda,\Lambda)^3}d\mathbf{q}\theta(q^2-\Lambda^2)\frac{k^{2n}}{q^{2(n+1)}},
\end{equation}
where $k^2=mE/\hbar^2$. The integral of the $n$-th term in the series is bounded as
\begin{equation}
0\le \int_{\mathbf{q}\in [-\Lambda,\Lambda)^3}d\mathbf{q}\theta(q^2-\Lambda^2)\frac{1}{q^{2(n+1)}}\le \frac{4\pi}{(2n-1)\Lambda^{2n-1}}.
\end{equation} 
Therefore, the following bound holds
\begin{equation}
0\le \int_{\mathbf{q}\in [-\Lambda,\Lambda)^3}d\mathbf{q}\frac{\theta(q^2-\Lambda^2)}{E-\hbar^2q^2/m}\le 4\pi\Lambda \sum_{n=1}^{\infty} \left(\frac{k}{\Lambda}\right)^{2n} = 4\pi \Lambda \frac{(k/\Lambda)^2}{1-(k/\Lambda)^2}.
\end{equation}
In the limit $\Lambda\to \infty$, the integral is bounded by zero from above and below, and it therefore converges to zero. Consequently, $I_0(E) = I_0(0)$ and the renormalised $R$-matrix in this regularisation scheme, for finite energies, is constant and identical to that obtained by means of spherically symmetric regularisation.

There is a subtlety that needs to be considered in order to obtain meaningful results. The integer cutoff in lattice sums with $L_s$ lattice sites (which we consider odd for convenience) is $\lambda=(L_s-1)/2$, which implies that the lattice site number-dependent real cutoff $\Lambda(L_s)$ is given by
\begin{equation}
\Lambda(L_s) = \frac{\pi }{d}\left(1-\frac{1}{L_s}\right).\label{Cutoff}
\end{equation}
The interaction {\it must} be renormalised by using the $L\to \infty$ limit of the cutoff, $\Lambda(\infty)$, since the limit $L\to \infty$ is taken at the start of the calculation when doing scattering theory, that is, the renormalisation prescription reads 
\begin{equation}
\frac{1}{g}=\frac{1}{g_R}-\frac{1}{(2\pi)^3}\frac{\pi b}{d}\frac{m}{\hbar^2}=\frac{1}{g_R}-\frac{m/\hbar^2}{(2\pi)^3}\Lambda(\infty),\label{Renorm2}
\end{equation}
where $b = -15.3482484734163\ldots$. Obviously, in the limit $L\to \infty$, we recover the result from minimal subtraction. 

\subsection{L{\"u}scher's formula}
We place the system in a cubic box of side length $L$ with periodic boundary conditions, at vanishing pair momentum $\mathbf{K}=0$. We will consider here the lowest unbound state. From BG theory, Eq. (\ref{discreteRmatrix}) and Eq. (\ref{BG3}), and setting $E=\epsilon(0)+\Delta E=\Delta E$, we obtain
\begin{equation}
\Delta E  = \frac{g}{L^3} + \frac{g}{L^3}\Delta E \sum_{\mathbf{q}\ne 0}\frac{1}{\Delta E - \epsilon(\mathbf{q})},\label{formula1}
\end{equation}
where the sum is assumed to be regularised and is renormalised to any given order by using the renormalisation prescription for the bare coupling constant $g$, Eq.~(\ref{Renorm2}). To next-to-leading order in $g/L^3$, the energy shift reads
\begin{equation}
\Delta E \approx \frac{g}{L^3}-\left(\frac{g}{L^3}\right)^2S_1(L),\label{DeltaE1}
\end{equation}
where we have defined 
\begin{equation}
S_n(L)=\sum_{\mathbf{q}\ne 0} \frac{1}{\left[\epsilon(\mathbf{q})\right]^n}.\label{Sn}
\end{equation}
In the above sums, $S_1$ is regularised with a cutoff $\Lambda(L_s)$ (given by Eq.~(\ref{Cutoff}), and will be seen to be renormalised by subtracting $I_0(0)$), calculated using the cutoff $\Lambda(\infty)$, while for $n>1$ the sums $S_n$ are regular and convergent. If we write
\begin{equation}
\frac{S_1(L)}{L^3}=\frac{m}{(2\pi)^3\hbar^2}\left(\int_{[-\Lambda(L),\Lambda(L)]^3}d\mathbf{q}\frac{1}{q^2}+\frac{\delta_1(L)}{L}\right),
\end{equation}
we obtain, to lowest order
\begin{equation}
\frac{S_1(L)}{L^3}\approx \frac{m}{\hbar^2(2\pi)^2}\left(b \Lambda(L) + \frac{\sigma}{L}\right),\label{S11}
\end{equation}
where $\sigma\approx -1.23951$. Upon renormalisation, we obtain
\begin{equation}
\Delta E \approx  \frac{4\pi \hbar^2}{m}\frac{a}{L^3}\left[1+2.837297 \frac{a}{L}\right].\label{DeltaE22}
\end{equation}
The above Equation (\ref{DeltaE22}) is in perfect agreement with L{\"u}scher's formula \cite{Luscher}, and emphasizes the importance of the subtlety regarding the renormalisation of the effective interaction in the naive continuum limit, as discussed in the previous section. 

\begin{figure}[t]
\includegraphics[width=1\textwidth]{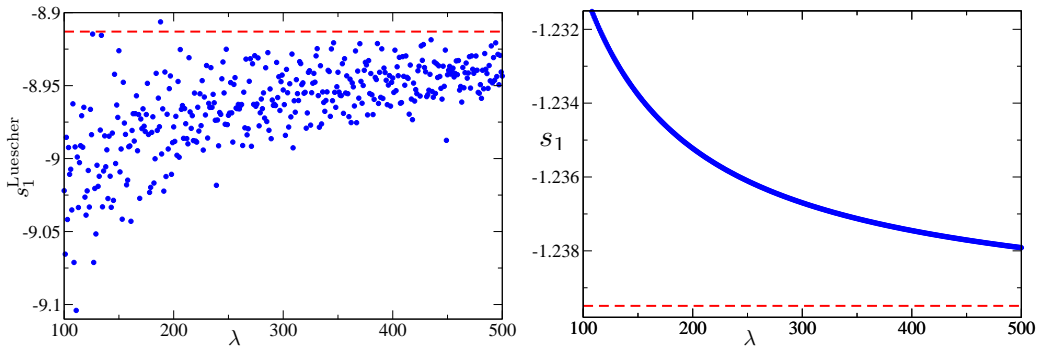}
\caption{Left: Blue dots are the numerically computed sums $s_1^{\mathrm{Luescher}}$ as a function of the integer cutoff $\lambda$ (see Eq. (\ref{s1Luscher})), and the dashed red line is its asymptotic value ($\lambda\to \infty$) quoted in refs. \cite{Luscher,Beane2004}. Right: Blue dots are the numerically computed sums $s_1$ as a function of $\lambda$ (see Eq. (\ref{s1Here})), and the dashed red line is the extrapolated value for $\lambda\to \infty$.}
\label{fig:Luscher1}
\end{figure}

There are several advantages of working with Eq. (\ref{DeltaE22}) with respect to the way one obtains L{\"u}scher's formula in the continuum limit {\it of a cubic lattice} \cite{Luscher}. Firstly, there is no need to project the interaction and the $R$-matrix onto their $s$-wave components, and since Eq. (\ref{DeltaE22}) only depends on the scattering length, this can be extracted with ease. Secondly, "brute force" summations (i.e. direct computation of the sums, equivalent to exact diagonalisation) provide smooth convergence for all the sums $S_n(L)$ in Eq. (\ref{Sn}), and equally for the energy-dependent sums in Eq.(\ref{formula1}). This is to be contrasted with the direct computation of the sum analogous to $S_1(L)$ in the $s$-wave channel \cite{Luscher,Beane2004}. In that case, the sums are restricted to $q<\Lambda$, while in the present case these are restricted to $|q_{\alpha}|<\Lambda$ ($\alpha=x,y,z)$. In Fig. \ref{fig:Luscher1} we plot the value of the sums
\begin{align}
s_1(\lambda)&=\sum_{\mathbf{n}\ne 0} \frac{\theta(\lambda,\mathbf{n})}{|\mathbf{n}|^2}-b\lambda,\label{s1Here}\\
s_1^{\mathrm{Luescher}}(\lambda)&=\sum_{\mathbf{n}\ne 0}\frac{\theta(\lambda^2-|\mathbf{n}|^2)}{|\mathbf{n}|^2},\label{s1Luscher}
\end{align}
as functions of the cutoff $\lambda$, where we have subtracted $\lambda = \lambda(L) = \lambda(\infty)-1/2$, while the second sum is renormalised with the spherically-symmetric prescription of ref. \cite{Beane2004}. Notice that the standard result for L{\"u}scher's sum is obtained here as $-1.23951-b/2 = -8.91363\ldots$. As is clearly observed in Fig. \ref{fig:Luscher1}, direct computation of the sums using projections of the interactions onto the $s$-wave channel gives rise to potentially high numerical errors. These would be inevitable in any computation relying on exact diagonalisation techniques, and would result in poor estimates of the scattering length. 

\subsection{Unitary limit}
In the unitary limit, i.e. for $a\to \pm \infty$, Eq. (\ref{DeltaE22}) is obviously incorrect. In the spherically symmetric case, Beane {\it et al.} \cite{Beane2004} used pionless effective field theory to derive the finite-size correction to the ground state energy when the scattering length is unnaturally large. This corresponds to the limit $a/L^3\to \pm \infty$. Here, we derive the corresponding expressions for the continuum limit of the lattice model in two different cases: (i) Beane {\it et al.}'s case ($a/L^3\to \pm \infty$) and (ii) the more natural case of $a=\mathcal{O}(L)$ in the limit $L\to \infty$.

For unnaturally large scattering lengths, Eq. (\ref{formula1}) gives
\begin{equation}
\Delta E = -\left[\sum_{\mathbf{q}\ne 0}\frac{1}{\Delta E-\epsilon(\mathbf{q})}\right]^{-1}.
\end{equation}
Clearly, we have $\Delta E = \mathcal{O}(1/L^2)$ in this case, and therefore $\Delta E$ is of the same order as the dispersion $\epsilon(\mathbf{q})$ at low momenta. The energy shift in this case is given by
\begin{equation}
\Delta E \approx -\frac{4\pi^2 \hbar^2}{mL^2}\left[0.09590\ldots + \mathcal{O}(1/aL)\right].\label{unnaturalLength}
\end{equation}
Again, the leading order term is identical to the one obtained in the spherically symmetric case \cite{Beane2004}.

In the case of large scattering lengths that scale as the length of the system, $a=\alpha L$, with $\alpha=\mathcal{O}(1)$, Eq. (\ref{formula1}) gives
\begin{equation}
\Delta E = \frac{4\pi\hbar^2\alpha}{mL^2}\frac{1}{1-\frac{4\pi\hbar^2\alpha}{mL^2}\sum_{\mathbf{q}\ne 0} [\Delta E-\epsilon(\mathbf{q})]^{-1}}.\label{caseii}
\end{equation}
Once more, $\Delta E = \mathcal{O}(1/L^2)$, and by setting $\Delta E \equiv (2\pi)^2\beta /L^2$,  Eq. (\ref{caseii}) becomes
\begin{equation}
\beta = \frac{\alpha}{\pi}\frac{1}{1-(\alpha/\pi)\sum_{\mathbf{n}\ne 0}(\beta-n^2)^{-1}}.\label{caseii2}
\end{equation}
The weak-coupling expansion in this case has the form
\begin{equation}
\beta \approx \frac{\alpha}{\pi}\left[1-2.837297\alpha\right].\label{beta}
\end{equation} 

Comparing Eqs. (\ref{unnaturalLength}) and (\ref{beta}), it is easy to observe that it is not always possible to distinguish the cases (i) and (ii) referred to above, corresponding to unnaturally large and "naturally" large scattering lengths, respectively, based on only one box size in the lowest energy shift. It can be tested, however, by choosing two different lattice sizes without changing the interaction parameters.

\section{Lattice case}
In the previous section we introduced the analog of L{\"u}scher's formula in the weak-coupling \cite{Luscher} and unitary \cite{Beane2004} limits for the na{\"i}ve continuum limit of the Hubbard model. In this section we study these expressions in the Hubbard model itself. 

\subsection{Scattering theory at low energy}
The Hubbard model, Eq. (\ref{HubbardHamiltonian}), is a one-parameter theory, since the physics of the model only depends on the ratio $U/J$. At zero energy, its $R$-matrix gives the scattering length. It does, however, have non-zero shape parameters and, of course, higher (even) partial waves, too. On the other hand, since it only depends on $U/J$ all shape parameters depend exclusively on the scattering length \footnote{This is analogous to what happens with a hard-sphere interaction \cite{Huang}.}, and so do the energies. In this way, scattering lengths can be extracted from energy shifts in finite lattices.

The on-shell zero-energy $R$-matrix is trivially shown to take the value
\begin{equation}
R=\frac{1}{1/U-W(0)},
\end{equation}
where 
\begin{equation}
W(E)=\frac{1}{(2\pi)^3}\mathcal{P}\int_{\mathrm{BZ}} d\mathbf{q} \frac{1}{E-\epsilon_0(\mathbf{q})}.
\end{equation}
Above, the energy dispersion becomes that of the Hubbard model, Eq. (\ref{dispersionHubbard}), for the relative motion of two particles at zero total momentum, i.e. 
\begin{equation}
\epsilon_0(\mathbf{q})=-4J\sum_{\alpha=x,y,z} \left[\cos(q_{\alpha})-1\right].
\end{equation}
The zero-energy integral is known exactly and was computed by Watson \cite{Watson}. It has the numerical value $JW(0)=-0.12636655\ldots\equiv \omega_0$. Therefore, the scattering length is given by 
\begin{equation}
\frac{m}{\hbar^2 a} = 4\pi \left[\frac{1}{U}-W(0)\right],\label{scatteringlengthU}
\end{equation}
where $m$ is the effective mass of the particles, related to the tunnelling rate by $J=\hbar^2/2m$ ($d=1$).

\subsection{Energy shifts}
The energy shift of the lowest unbound state is readily obtained from BG theory, Eq. (\ref{discreteRmatrix}) and Eq. (\ref{BG3}), and reads in this case
\begin{equation}
\Delta E = \frac{U/L^3}{1-\frac{U}{L^3}\sum_{\mathbf{q}\ne 0} \frac{1}{\Delta E -\epsilon_0(\mathbf{q})}}.\label{DeltaEU}
\end{equation}
In terms of the scattering length, Eq. (\ref{scatteringlengthU}), Eq. (\ref{DeltaEU}) becomes
\begin{equation}
\Delta E = \frac{4\pi \hbar^2a}{mL^3} \left[1+4\pi a(2\omega_0) - \frac{4\pi \hbar^2 a}{mL^3}\sum_{\mathbf{q}\ne 0}\frac{1}{\Delta E - \epsilon_0(\mathbf{q})}\right]^{-1}. 
\end{equation}
In the weak-coupling limit, we therefore obtain
\begin{equation}
\Delta E \approx \frac{4\pi \hbar^2 a}{mL^3}\left[1-4\pi a (2\omega_0)+\frac{4\pi \hbar^2 a}{mL^3}\sum_{\mathbf{q}\ne 0}\frac{1}{\Delta E - \epsilon_0(\mathbf{q})}\right] .\label{LuscherLattice1}
\end{equation}
The sum on the right hand side of Eq. (\ref{LuscherLattice1}) can be estimated as follows. Consider a cubic region $\Omega=[-\eta,\eta]^3-\{(0,0,0)\}$, with $\eta>0$ being much smaller than the Brillouin zone's extent ($\eta\llt \pi$). The sum can be split into the two regions $BZ-\Omega$ and $\Omega$. The sum over $BZ-\Omega$ can be approximated by an integral, and therefore
\begin{equation}
\frac{1}{L^3} \sum_{\mathbf{q}\ne 0}\frac{1}{\epsilon_0(\mathbf{q})} \approx \frac{1}{(2\pi)^3} \int_{BZ-\Omega} d\mathbf{q} \frac{1}{\epsilon_0(\mathbf{q})} +\frac{1}{L^3} \sum_{\mathbf{q}\in \Omega} \frac{1}{\epsilon_0(\mathbf{q})}.
\end{equation}
In the region $\Omega$, the full lattice dispersion can be approximated by
\begin{equation}
\epsilon_0(\mathbf{q})\approx 2J \mathbf{q}^2= \left(\frac{2\pi}{L}\right)^2 2J |\mathbf{n}|^2.
\end{equation}
setting $\eta=(2\pi \bar{N}/L)$, we obtain in the limit $\bar{N}\to \infty$ with $\bar{N}/L \to 0$
\begin{equation}
\sum_{\mathbf{q}\in \Omega} \frac{1}{\epsilon_0(\mathbf{q})}\approx \left(\frac{L}{2\pi}\right)^2 \frac{1}{2J} \sum_{\mathbf{n}\ne 0} \frac{\theta(\bar{N},\mathbf{n})}{|\mathbf{n}|^2} = \left(\frac{L}{2\pi}\right)^2 \frac{1}{2J} \left[b\bar{N} + s_1\right],
\end{equation}
where $b=15.3482484734169\ldots$ and $s_1=-1.23945\ldots$, as calculated in the previous section. The integral in the region $\mathrm{BZ}-\Omega$ is easily calculated as
\begin{equation}
\int_{\mathrm{BZ}-\Omega}d\mathbf{q} \frac{1}{\epsilon_0(\mathbf{q})}\approx \int_{\mathbf{BZ}}d\mathbf{q}\frac{1}{\epsilon_0(\mathbf{q})} - \frac{1}{2J}\int_{\Omega} d\mathbf{q}\frac{1}{q^2} = \int_{\mathbf{BZ}}\frac{1}{\epsilon_0(\mathbf{q})}-\left(\frac{L}{2\pi}\right)^2 \frac{b(\bar{N}+1/2)}{2J}.
\end{equation}
Using the above results in Eq. (\ref{LuscherLattice1}), the energy shift becomes
\begin{equation}
\Delta E \approx \frac{4\pi \hbar^2}{m} \frac{a}{L^3} \left[1+2.837297\frac{a}{L}\right].\label{DeltaE2Lattice}
\end{equation}
The above expression coincides with the corresponding formula in the na{\"i}ve continuum limit, Eq. (\ref{DeltaE22}), as expected. However, Eq. (\ref{DeltaE2Lattice}) has been obtained without assuming the continuum limit at all energies, and it proves that the na{\"i}ve continuum limit -- with its associated subtleties -- provides an adequate description of low-energy scattering in the weak-coupling limit.

The corresponding expressions in the unitary limits (both the natural and unnatural cases) are obtained in a fashion completely analogous to the one leading to Eq. (\ref{DeltaE2Lattice}) and are identical to the expressions obtained using the na{\"i}ve continuum limit in the previous section.

\section{Conclusions and outlook}
Besides purely theoretical implications, the results reported in this work have interesting applications. For instance, two-particle scattering in a realistic three-dimensional lattice is of great relevance for condensed matter and cold atomic systems in optical lattices, as it is necessary in order to renormalise the coupling constant ($U/J$) in the Hubbard model. Unfortunately, the calculation of low-energy scattering properties of this system via direct use of collision theory is generally very involved, the reason being the genuine multi-channel structure of these systems. One way to estimate the scattering properties is to solve for the ground state energy of the realistic model consisting of only a few lattice wells (the number of wells being $L_s^3$). With the formulas derived in this article, one can fit the scattering length, and therefore the interaction coupling constant, to match the energy shift in the realistic calculation in a small box. Generalisations of the current work to study, e.g. effective interparticle interactions in graphene \cite{GrapheneReview} and graphene-like optical lattices \cite{GrapheneOL}, are possible by using the methods presented here. In intrinsic graphene, as opposed to the cold atomic case and extrinsic graphene \cite{Hofmann}, however, the inclusion of the Coulomb interaction is necessary and the problem is more complicated, but can be done by extending the methods for continuous theories \cite{BeaneCoulomb}. In particular, the rigorous construction of the continuum quasi-relativistic (it is not Lorentz invariant) effective field theory around the Dirac points is still an open problem \footnote{The two-body problem in the na{\"i}ve continuum limit has been studied in, e.g. ref. \cite{Zapata}. There, however, the interaction is purely phenomenological.}.

It will be interesting to study the so-called Bertsch parameter \cite{Baker}, defined as the ratio between the ground-state energy of a resonantly interacting three-dimensional Fermi gas in the s-wave channel to the non-interacting ground state energy, but for the Hubbard model itself. This can be done using quantum Monte Carlo simulations \cite{Endres,Drut}. Given that the energy shift, considered here, for two fermions at the lattice resonance is the same as that obtained from the pure s-wave results, it is to be expected that the lattice version of Bertsch parameter will be identical, or similar, with its continuum, pure s-wave counterpart. Its calculation would actually be much less involved since the elimination of higher-partial waves and effective range, cleverly done in ref. \cite{Endres}, is not necessary at all.

\section*{Acknowledgements}
We thank L.~G. Phillips for illuminating discussions. The authors acknowledges support from EPSRC grant No. EP/J001392/1 
and from the Danish Council for Independent Research under the Sapere Aude program.

\bibliographystyle{unsrt}

\begin{thebibliography}{00}
\bibitem{Huang}
K. Huang and C.~N. Yang, Phys. Rev. {\bf 105}, 767 (1957).

\bibitem{Fermi}
E. Fermi, Ricerca Sci. {\bf 7}, 13 (1936).

\bibitem{NuclearForces}
E. Epelbaum, H.~-W. Hammer and U.~-G. Meissner, Rev. Mod. Phys. {\bf 81}, 1773 (2009).

\bibitem{Luscher}
M. L{\"u}scher, Commun. Math. Phys. {\bf 105}, 153 (1986).

\bibitem{BeanePDS}
S. Weinberg, Phys. Lett. B {\bf 251}, 288 (1990); D.~R. Phillips, S.~R. Beane and T.~D. Cohen, Nucl. Phys. A {\bf 631}, 447 (1998); D.~B. Kaplan, M.~J. Savage and M.~B. Wise, Phys. Lett. B {\bf 424}, 390 (1998); Nucl. Phys. B {\bf 534}, 329 (1998).

\bibitem{Beane2004}
S.~R. Beane, P.~F. Bedaque, A. Parre{\~n}o and M.~J. Savage,
Phys. Lett. B {\bf 585}, 106 (2004).

\bibitem{BeaneQCD}
S.~R. Beane, P.~F. Bedaque, K.~O. Orginos and M.~J. Savage,
Phys. Rev. Lett. {\bf 97}, 012001 (2006).

\bibitem{ValienteLattices}
M. Valiente, Phys. Rev. A {\bf 81}, 042102 (2010); M. Valiente and D. Petrosyan, J. Phys. B: At. Mol. Opt. Phys. {\bf 42}, 121001 (2009); {\it ibid.} {\bf 41}, 161002 (2008); M. Valiente, M. K{\"u}ster and A. Saenz, EPL {\bf 92}, 10001 (2010); M. Valiente and K. M{\o}lmer, Phys. Rev. A {\bf 84}, 053628 (2011); 

\bibitem{FetterWalecka}
A.~L. Fetter and J.~D. Walecka, {\it Quantum Theory of Many-Particle Systems} (Dover, N.Y., 2003).

\bibitem{Tan}
S. Tan, Ann. Phys. (N.Y.) {\bf 323}, 2952 (2008).

\bibitem{ValienteTan}
M. Valiente, Phys. Rev. A {\bf 85}, 014701 (2012).

\bibitem{Watson}
G.~N. Watson, Quart. J. Math., Oxford Ser. 2 {\bf 10}, 266 (1939).

\bibitem{GrapheneReview}
A.~H. Castro Neto, F. Guinea, N.~M.~R. Peres, K.~S. Novoselov and A.~K. Geim,
Rev. Mod. Phys. {\bf 81}, 109 (2009).

\bibitem{GrapheneOL}
L. Tarruell, D. Greif, T. Uehlinger, G. Jotzu and T. Esslinger,
Nature {\bf 483}, 303 (2012).

\bibitem{Hofmann}
J. Hofmann, E. Barnes and S. Das Sarma,
Phys. Rev. Lett. {\bf 113}, 105502 (2014).

\bibitem{BeaneCoulomb}
S.~R. Beane and M.~J. Savage,
Phys. Rev. D {\bf 90}, 074511 (2014).

\bibitem{Zapata}
C. Gaul, F. Dom{\'i}nguez-Adame, F. Sols and I. Zapata,
Phys. Rev. B {\bf 89}, 045420 (2014).

\bibitem{Baker}
G.~A. Baker, Phys. Rev. C {\bf 60}, 05311 (1999).

\bibitem{Endres}
M~G. Endres, D.~B. Kaplan, J.-W. Lee and A. Nicholson,
Phys. Rev. A {\bf 87}, 023615 (2013).

\bibitem{Drut}
J.~E. Drut, Phys. Rev. A {\bf 86}, 013604 (2012).

\end{thebibliography}

\end{document}